\documentclass[pdftex,12pt,letter]{article} 
\usepackage{./mylay}

\usepackage{amsthm}
\usepackage{amsmath,amsfonts, amssymb}
\usepackage{authblk}

\pdfoutput=1

\def\mc{\mathcal}

\newcommand{\E}{{\mathbb E}}

\usepackage{xspace}

\newcommand{\pcrp}[1]{q_{\theta}^{\mathrm{CRP}}(#1)}
\newcommand{\pcrpmix}[1]{\tilde{q}_{m,n}^{\mathrm{CRP}}(#1)}

\newcommand{\ppy}[1]{q_{\alpha,\theta}^{\mathrm{PY}}(#1)}

\providecommand{\eg}{{\emph{e.g.\ }}}

\newcommand{\ptr}{{\psi}}

\newcommand{\Ptr}{{\Psi}}
\newcommand{\Ptrs}[1]{{\Ptr^{#1}}}
\newcommand{\Ptrn}{{\Ptrs n}}

\def\btheta{\bar{\theta}}
\def\pcrpmixij{\tilde{q}_{i,j}^{\mathrm{CRP}}}

\def \cI     {{\cal I}}
\def \cO     {{\cal O}} 
\def \cP     {{\cal P}} 
\def \upto  {{,}\ldots{,}}
\def \Paren#1{{\left({#1}\right)}}

\def \sets#1{{\{#1\}}}

\def \skpbld#1{\par\noindent\textbf{#1}\quad}

\def \Proof    {\skpbld{Proof}}

\renewcommand\footnotemark{}

\providecommand{\keywords}[1]{\textbf{\textit{Keywords---}} #1}

\title{Redundancy of Exchangeable Estimators } 

\author{ Narayana P. Santhanam$^{1}$, Anand D. Sarwate$^{2}$ and Jae Oh Woo$^{3}$ } 

\thanks{$^{1}$Department of Electrical Engineering, University of Hawaii at Manoa, 2540 Dole Street, Honolulu, HI 96822, USA, \texttt{nsanthan@hawaii.edu}}
\thanks{$^{2}$Department of Electrical and Computer Engineering, Rutgers, The State University of New Jersey, 94 Brett Road, Piscataway, NJ 08854 , USA, \texttt{asarwate@ece.rutgers.edu}}
\thanks{$^{3}$Applied Mathematics Program, Yale University, 51 Prospect St, New Haven, CT 06511, USA, \texttt{jaeoh.woo@yale.edu}}

\begin{document}
\maketitle

\abstract{
Exchangeable random partition processes are the basis for
Bayesian approaches to statistical inference in large alphabet settings.
On the other hand, the notion of the \emph{pattern} of a sequence
provides an information-theoretic framework for data compression in large alphabet scenarios.
Because data compression and parameter
estimation are intimately related, we study the
redundancy of Bayes estimators coming from Poisson-Dirichlet priors
(or ``Chinese restaurant processes'') and the Pitman-Yor prior.
This provides an understanding of these estimators in the setting of unknown discrete alphabets
from the perspective of universal compression.
In particular, we identify relations between alphabet sizes
and sample sizes where the redundancy is small, thereby
characterizing useful regimes for these estimators.}


\section{Introduction}

A number of statistical inference problems of significant contemporary interest, 
such as text classification, language modeling and DNA microarray analysis, 
require inferences based on observed 
sequences of symbols in which the sequence length or sample size
is comparable or even smaller than the set of symbols, the alphabet.
For instance, language models for speech recognition estimate
distributions over English words using text examples much smaller than
the vocabulary. 

Inference in this setting has received a lot of 
attention, from Laplace~\cite{Lap:pe,DeMor38,DeMor45} in the 18th century, 
to~Good~\cite{Goo53} in the mid-20th century, to an explosion of work in the
statistics~\cite{BM73,Kin75:pd,Kin78,Kin78:de,DF80,Ald85,Zab92,Pit95,PY97},
information theory \cite{CB90,CB94,OSZ03,OSZ03:s+f,Rya08,WVK11} and
machine learning~\cite{Nad91,GC94,MS00,DM04} communities in the last few decades.
A major strand in the information theory literature on the subject has
been based on the notion of patterns. The pattern of a sequence characterizes the repeat structure in the sequence, which is the information that can be described well (see Orlitsky {\em et al.}~\cite{OSVZ04} for formal
characterizations of this idea). The statistical
literature has emphasized the importance of exchangeability, which generalizes the notion of independence.

We consider measures over infinite sequences $X_1,X_2\ldots$, where
$X_i$ come from a countable (infinite) set (the alphabet). Let
$\cI$ be the collection of all distributions over countable
(potentially infinite) alphabets. For $p \in \cI$, let $p^{(n)}$
denote the product distribution corresponding to an independent and
identically distributed (i.i.d.) sample $X_1^n = (X_1, X_2, \ldots,
X_n)$, where $X_i \sim p$. Let $\cI^{(n)}$ be the collection of all
such i.i.d. distributions on length $n$ sequences drawn from countable
alphabets. Let $\cI^\infty$ be the collection of all measures over
infinite sequences of symbols from countable alphabets $X_1,X_2\ldots$,
where the $\{X_i\}$ are i.i.d. according to some distribution in
$\cI$. The measures are constructed by extending to the Borel sigma
algebra the i.i.d. probability assignments on finite length sequences, namely
$\cI^{(n)}$, $n\ge1$. We call $\cI^\infty$ the set of i.i.d. measures.

Based on a sample $X_1^n = (X_1, X_2, \ldots, X_n)$ from an unknown
$p^{(n)} \in \cI^{(n)}$ (or equivalently, the corresponding measure in
$\cI^\infty$), we want to create an estimator $q_n$, which
assigns probabilities to length-$n$ sequences. We are interested in
the behavior of the sequence of estimators $\{q_n : n = 1, 2, \ldots
\}$. With some abuse of notation, we will use $q$ to denote the
estimator $q_n$ when the sample size $n$ is clear from context. We
want $q_n$ to approximate $p^{(n)}$ well; in particular, we would like
$q_n$ to neither overestimate nor underestimate the probability of
sequences of length $n$ under the true $p^{(n)}$.

Suppose that there exist $R_n > 0$ and $A_n > 0$, such that
for each $p \in \cI$, we have:
\[
p^{(n)}\Paren{ \left\{ X_1^n: q_n(X_1^n) > R_n p^{(n)}(X_1^n) \right\} } < 1/A_n.
\]
If $A_n$ is any function of $n$ that grows sufficiently fast with $n$,
the sequence $\{q_n\}$ does not asymptotically overestimate probabilities of length-$n$
sequences by a factor larger than than $R_n$ with probability one, no matter what 
measure $p \in \cI$ generated the sequences. 

Protecting against underestimation is not so simple. 
The
redundancy of an estimator $q_n$ (defined formally in
Section~\ref{sec:redundancy}) for a length $n$ sequence $x_1^n$ measures how
closely $q_n(x_1^n)$ matches: 
\[
\max_{p^{(n)} \in \cI^{(n)}} p^{(n)}(x_1^n),
\] 
the largest probability assigned to $x_1^n$ by any distribution in
$\cI^n$. The estimator redundancy usually either maximizes the
redundancy of a sequence or takes the expectation over all
sequences. Ideally, we want the estimator redundancy to grow
sublinearly in the sequence length $n$, so that the per-sample redundancy
vanishes as $n \to \infty$. If so, we call the estimator
universal for $\cI$. Redundancy thus captures how well $q$
performs against the collection $\cI$, but the connections between
estimation problems and compression run deeper.

In this paper, we consider estimators formed by taking a measure 
(prior) on $\cI$. Different priors induce
different distributions on the data $X_1^n$. We think of the prior as
randomly choosing a distribution $p$ in $\cI$, and the observed data $X_1^n$ is generated according to this $p$. How much information about the underlying
distribution $p$ can we obtain from the data (assuming we know the prior)?
Indeed, a well known result~\cite{Gal76,DL80,Rya79} proves that the redundancy of the
best possible estimator for $\cI^\infty$ equals the maximum (over all choices of priors) 
information (in bits) that is present about the underlying source in a length $n$
sequence generated in this manner.

Redundancy is well defined for finite alphabets; recent
work~\cite{OSZ03} has formalized a similar framework for countably
infinite alphabets. This framework is based on the notion of patterns
of sequences that abstract the identities of symbols and indicate only
the relative order of appearance. For example, the pattern of FEDERER
is 1232424, while that of PATTERN is 1233456. The crux of the idea is
that instead of considering the set of measures $\cI^{\infty}$ over
infinite sequences, we consider the set of measures induced over
patterns of the sequences. It then follows that now our estimate
$q_{{}_\Psi}$ is a measure over patterns. While the variables in the
sequence are i.i.d., the corresponding pattern merely corresponds to a
exchangeable random partition.  We can associate a predictive distribution with the
pattern probability estimator $q_\Psi$.  This is an estimate of the
distribution of $X_{n+1}$ given the previous observations, and it
assigns probabilities to the event that $X_{n+1}$ will be ``new'' (has
not appeared in $X_1^n$) and probabilities to the events that
$X_{n+1}$ takes on one of the values that has been seen so far.

The above view of estimation also appears in the statistical literature on
Bayesian nonparametrics that focuses on exchangeability. Kingman
\cite{Kin80} advocated the use of exchangeable random partitions
to accommodate the analysis of data from an alphabet that is not
bounded or known in advance. A more detailed discussion of the
history and philosophy of this problem can be found in the works of
Zabell \mbox{\cite{Zab92,Zab97}} collected in \cite{Zab05}. One of the most
popular exchangeable random partition processes is the ``Chinese
restaurant process'' \cite{Ald85}, which is a special case of the
Poisson--Dirichlet or Pitman--Yor process \cite{Pit96b,PY97}. These
processes can be viewed as prior distributions on the set of all
discrete distributions that can be used as the basis for estimating
probabilities and computing predictive~distributions.

In this paper, we evaluate the performance of the sequential
estimators corresponding to these exchangeable partition processes. 
As described before, $\cI$ is the collection of all
distributions over countable (potentially infinite) alphabets, 
and $\cI^\infty$ is the collection of all i.i.d. measures with single letter
marginals in $\cI$. Let $\cI_\Psi$ be the collection of all measures
over patterns induced by measures in $\cI^\infty$. We evaluate the
redundancy of estimators based on the Chinese restaurant process (CRP), 
the Pitman--Yor (PY) process and the Ewen's sampling formula against $\cI_\Psi$.
In the context of sequential estimation, early work~\cite{OSZ03} showed that for
the collection $\cI_\Psi$ of measures over patterns, universal estimators do exist: the normalized
redundancy is $O(n^{1/2})$.  More recent work \cite{AcharyaDJOS:13isit,AcharyaDO:12nips} proved tight bounds on worst-case and average redundancy; these results show that there are sequential estimators with normalized redundancy of $O(n^{1/3})$.  However, these estimators are computationally intensive and (generally speaking) infeasible in practice.  Acharya et al.~\cite{AcharyaJOS:13colt} demonstrated a linear-time estimator with average redundancy $O(n^{1/2})$, improving over the earlier constructions achieving $O(n^{2/3})$~\cite{OSZ03}.
By contrast, estimators such as the CRP or the PY estimator were not
developed in a universal compression framework, but have been very successful from a practical standpoint. The goal of this paper is to understand these Bayesian estimators from the universal compression perspective.

For the case of the estimators studied in nonparametric Bayesian
statistics, our results show that they are in general neither weakly
nor strongly universal when compressing patterns or equivalently
exchangeable random partitions. While the notion of redundancy is in
some sense different from other measures of accuracy, such as
the concentration of the posterior distribution about the true
distribution, the parameters of the CRPs or the PY processes that do
compress well often correspond to the maximum likelihood estimates
obtained from the sample.

Because we choose to measure redundancy in the worst case over $p$,
the underlying alphabet size may be arbitrarily large with respect to
the sample size $n$. Consequently, for a fixed sample of size $n$, the
number of symbols could be large, for example a constant fraction of
$n$. The CRP and PY estimators do not have good redundancy against
such samples, since they are not the cases the estimators are designed
for. However, we can show that a mixture of estimators corresponding
to CRP estimators is weakly universal. This mixture is made by
optimizing individual CRP estimators that (implicitly) assume a bound
on the support of $p$. If such a bound is known in advance, we can
derive much tighter bounds on the redundancy. In this setting, the
two-parameter Poisson--Dirichlet (or Pitman--Yor) estimator is superior
to the estimator derived from the Chinese restaurant process.

In order to describe our results, we require a variety of definitions
from different research communities. In the next section, we
describe this preliminary material and place it in context before describing the main results in Section \ref{sec:results}.

\section{Preliminaries}
\label{sec:prelim}

In this paper, we use the ``big-$O$'' notation. A function $f(n) = O(g(n))$ if there exists a positive constant $C,$ such that for sufficiently large $n$, $|f(n)| \le C |g(n)|$. A function $f(n) =\Omega(g(n))$, if there exists a positive constant $C'$, such that for sufficiently large $n$, $|f(n)| \ge C' |g(n)|$. A function \mbox{$f(n) = \Theta(g(n))$}, if $f(n) = O(g(n))$ and $f(n) = \Omega(g(n))$.

Let $\mc{I}_k$ denote the set of all probability distributions on alphabets of size $k$ and $\mc{I}_{\infty}$ be all probability distributions on countably infinite alphabets, and let:
	\begin{align*}
	\mc{I} = \mc{I}_{\infty} \cup \bigcup_{k \ge 1} \mc{I}_k
	\end{align*}
be the set of all discrete distributions irrespective of support and support size.

For a fixed $p$, let $x_1^n = (x_1, x_2, \ldots, x_n)$ be a sequence drawn i.i.d. according to $p$. We denote the pattern of $x_1^n$ by $\ptr_1^n$. The pattern is formed by taking $\ptr_1 = 1$ and: 
	\begin{align*}
	\ptr_i = \left\{
	\begin{array}{ll}
	\ptr_{j} & x_i = x_j,\ j < i \\
	1 + \max_{j < i} \ptr_j & x_i \ne x_j,\ \forall j < i
	\end{array}
	\right.
	\end{align*}
For example, the pattern of $x_1^7 = \mathsf{FEDERER}$ is $\psi_1^7 = 1232424$. Let $\ptr^n$ be the set of all patterns of length~$n$. We write $p(\ptr^n)$ for the probability that a length-$n$ sequence generated by $p$ has
pattern $\ptr^n$. For a pattern $\ptr_1^n$, we write $\phi_{\mu}$ for the number of symbols that appear $\mu$ times in $\ptr_1^{n}$, and $m = \sum \phi_{\mu}$ is the number of distinct symbols in $\psi_1^n$. We call $\phi_{\mu}$ the prevalence of $\mu$. Thus, for $\mathsf{FEDERER}$, we have $\phi_1 = 2$, $\phi_2 = 1$, $\phi_3 = 1$ and $m = 4$.

\subsection{Exchangeable Partition Processes}

An exchangeable random partition refers to a sequence 
$(C_n:n\in\mathbb{N})$, where $C_n$ is a random partition of the
set $[n]=\{1,2,\ldots,n\}$, satisfying the following conditions:
(i) the probability that $C_n$ is a particular partition
depends only on the vector $(s_1, s_2, \ldots,s_n)$,
where $s_k$ is the number of parts in the partition of
size $k$; and (ii) the realizations of the sequence are
consistent in that all of the parts of $C_n$ are also parts of
the partition $C_{n+1}$, except that the new element $n+1$ 
may either be in a new part of $C_{n+1}$ by itself 
or has joined one of the existing parts of $C_n$.

For a sequence $X_1, \ldots, X_n$ from a discrete alphabet,
one can partition the set $[n]$ into component sets $\{A_x\}$, where
$A_x = \{i : X_i = x\}$ are the indices corresponding to the positions 
in which $x$ has appeared. When the $\{X_i\}$ are drawn i.i.d. from a distribution in $\mc{I}$, the corresponding sequence of random partitions
is called a paintbox process.

The remarkable Kingman representation theorem~\cite{Kin78:de} states that the probability measure induced by any
exchangeable random partition is a mixture of paintbox processes, where the mixture is taken using a probability measure
(``prior'' in Bayesian terminology) on the class of paintbox processes. Since each paintbox process corresponds to a
discrete probability measure (the one such that i.i.d. $X_i$ drawn from it produced the paintbox process), the prior may
be viewed as a distribution on the set of probability measures on a countable alphabet. 
For technical reasons, the alphabet is assumed to be hybrid, with a discrete part, as well as a continuous
part, and also, one needs to work with the space of ordered probability vectors (see~\cite{Pit} for details).

\subsection{Dirichlet Priors and Chinese Restaurant Processes}

Not surprisingly, special classes of priors give rise to special classes of
exchangeable random partitions. One particularly nice class of priors
on the set of probability measures on a countable alphabet is that of
the Poisson--Dirichlet priors~\cite{Fer73,BM73,RS07} 
(sometimes called Dirichlet processes, since they live on
the infinite-dimensional space of probability measures
and generalize the usual finite-dimensional Dirichlet distribution). 

The Chinese restaurant process (or CRP) is related to the so-called Griffiths--Engen--McCloskey (GEM) 
distribution with parameter $\theta$, denoted by $\mathsf{GEM}(\theta)$. 
Consider $W_1, W_2, \ldots$ drawn i.i.d. according to a $\mathsf{Beta}(1,\theta)$ distribution, and set:
	\begin{align*}
	p_1 &= W_1 \\
	p_i &= W_i \prod_{j < i} (1 - W_i) \qquad \forall i > 1
	\end{align*}
This can be interpreted as follows: take a stick of unit length and break it into pieces 
of size $W_1$ and $1 - W_1$. Now take the piece of size $1 - W_1$ and break off 
a $W_2$ fraction of that. Continue in this way. The resulting lengths of the sticks 
create a distribution on a countably infinite set. The distribution of the sequence 
$p = (p_1, p_2, \ldots)$ is the $\mathsf{GEM}(\theta)$ distribution.

\bRemark
Let $\pi$ denote the elements of $p$ sorted in decreasing order, so that 
$\pi_1 \ge \pi_2 \ge \cdots$. Then, the distribution of $\pi$ is the Poisson--Dirichlet 
distribution $\mathsf{PD}(\theta)$ as defined by Kingman.
\eRemark

Another popular class of distributions on probability vectors is the Pitman--Yor family
of distributions \cite{PY97}, also known as the two-parameter Poisson--Dirichlet 
family of distributions \linebreak $\mathsf{PD}(\alpha,\theta)$. The two parameters here are a discount parameter $\alpha \in [0,1]$ and a strength parameter $\theta > -\alpha$.	
The distribution $\mathsf{PD}(\alpha,\theta)$ can be generated in a similar way as the 
Poisson--Dirichlet distribution $\mathsf{PD}(\theta)=\mathsf{PD}(0,\theta)$ described earlier. Let each $W_i$ be drawn independently according to a \linebreak $\mathsf{Beta}(1 - \alpha, \theta + i \alpha)$ distribution, and again, set:
	\begin{align*}
	\tilde{p}_1 &= W_1 \\
	\tilde{p}_i &= W_i \prod_{j < i} (1 - W_i) \qquad \forall i > 1
	\end{align*}
A similar ``stick-breaking'' interpretation holds here, as well. Now, let $p$ be equal to the 
sequence $\tilde{p}$ sorted in descending order. The distribution of $p$ is $\mathsf{PD}(\alpha,\theta)$.
If we have $\alpha < 0$ and $\theta = r |\alpha|$ for integer $r$, we may obtain a symmetric Dirichlet distribution of dimension $r$.

\subsection{Pattern Probability Estimators}

Given a sample $x_1^n$ with pattern $\ptr^n$, we would like to produce a pattern probability estimator. This is a function of the form $q(\ptr_{n+1} | \ptr^n)$ that assigns a probability of seeing a symbol previously seen in $\ptr^n$, as well as a probability of seeing a new symbol. In this paper, we will investigate two different pattern probability estimators based on Bayesian models.

The Ewens sampling formula \cite{Ewe72,KM72,W74}, which has its origins in theoretical population genetics, is a formula for the probability mass function of a marginal of a CRP corresponding to a fixed population size. In other words, it specifies the probability of an exchangeable random partition of $[n]$ that is obtained when one uses the Poisson--Dirichlet $\mathsf{PD}(\theta)$ prior to mix paintbox processes. Because of the equivalence between patterns and exchangeable random partitions,
it estimates the probability of a pattern $\ptr_1^{n}$ via the following formula:
	\begin{equation}
	\pcrp{\ptr_1\upto \ptr_n} 
	= \frac{\theta^m}{\theta (\theta+1) \cdots (\theta + n - 1)} \prod_{\mu=1}^{n} [(\mu - 1)!]^{\phi_{\mu}}.
	\label{eq:CRPfull}
	\end{equation}
Recall that $\phi_{\mu}$ is the number of symbols that appear $\mu$ times in $\ptr_1^{n}$.
In particular, the predictive distribution associated to the Ewens sampling formula or Chinese restaurant process is: 
	\begin{align}
\pcrp{\ptr|\ptr_1\upto \ptr_n} 
=\begin{cases}
\frac{\mu}{n+\theta} & \ptr \text{ appeared $\mu$ times}\\
      & \text{ in $\ptr_1\upto\ptr_n$; }\\
\frac{\theta}{n+\theta} & \ptr \text{ is a new symbol.}
\end{cases}
	\label{eq:CRPstep}
	\end{align}

More generally, one can define the Pitman--Yor predictor (for $\alpha\in[0,1]$ and $\theta>-\alpha$) as:
	\begin{align}
	\ppy{\ptr|\ptr_1\upto \ptr_n}
=\begin{cases}
\frac{\mu - \alpha}{n+\theta} & \ptr \text{ appeared $\mu$ times}\\
      & \text{ in $\ptr_1\upto\ptr_n$; }\\
\frac{\theta + m \alpha}{n+\theta} & \ptr \text{ is a new symbol.}
\end{cases}
	\label{eq:PYstep}
	\end{align}
where $m$ is the number of distinct symbols in $\psi_1^n$. 
The probability assigned by the Pitman--Yor predictor to a pattern $\ptr_1^n$ is therefore:
	\begin{equation}
	\ppy{\ptr_1\upto \ptr_n} = 
	\frac{\theta (\theta + \alpha) (\theta + 2 \alpha) \cdots (\theta + (m-1) \alpha)}{\theta (\theta+1) \cdots (\theta + n - 1)} \prod_{\mu=1}^{n} \left( \frac{\Gamma(\mu - \alpha)}{\Gamma(1 - \alpha)} \right)^{\phi_{\mu}}.
	\label{eq:PYfull}
	\end{equation}
Note that $\Gamma(\mu - \alpha)/\Gamma(1 - \alpha) = (\mu - \alpha - 1)(\mu - \alpha - 2) \cdots (1 - \alpha)$.

\subsection{Strong Universality Measures: Worst-Case and Average }
 \label{sec:redundancy}

How should we measure the quality of a pattern probability predictor $q$? 
We investigate two criteria here: the worst-case and the average-case redundancy. 
The redundancy of $q$ on a given pattern $\ptr^n$ is:
	\begin{align*}
	R(q,\ptr^n)\ed 
	\sup_{p\in\cI} 
		\log \frac{p(\ptr^n)}{q(\ptr^n)},
	\end{align*}
The worst-case redundancy of $q$ is defined to be:
	\begin{align*}
	\hat{R}(q)\ed 
	\max_{ \ptr^n\in\Ptrn }\sup_{p\in\cI} 
		\log \frac{p(\ptr^n)}{q(\ptr^n)}=
	\sup_{p\in\cI} \max_{ \ptr^n\in\Ptrn }
		\log \frac{p(\ptr^n)}{q(\ptr^n)}
	\end{align*}
 Recall that $p(\ptr^n)$ just denotes the
 probability that a length-$n$ sequence generated by $p$ has pattern $\ptr^n$; it is unnecessary to specify the
 support here. Since the set of length-$n$ patterns is finite, there is no need for a supremum in the outer maximization
 above. The worst-case redundancy is often referred to as the per-sequence redundancy, as well.

The average-case redundancy replaces the $\max$ over patterns with an expectation over $p$:
	\begin{align*}
	\bar{R}(q) &\ed
	\sup_{p\in\cI} \mathbb{E}_p \left[ \log \frac{p(\ptr^n)}{q(\ptr^n)} \right] \\
	&= \sup_{p\in\cI} D\left( p ~\|~ q \right),
	\end{align*}
 where $D(\cdot \| \cdot)$ is the Kullback--Leibler divergence or relative
 entropy. That is, the average-case redundancy is nothing but the worst-case
 Kullback--Leibler divergence between the distribution $p$ and the predictor
 $q$.

A pattern probability estimator is considered ``good'' if the
worst-case or average-case redundancies are sublinear in
$n$ or $\hat{R}(q)/n \to 0$ and $\bar{R}(q)/n \to 0$ as $n \to \infty$. 
Succinctly put, redundancy that is sublinear in $n$
implies that the underlying probability of a sequence can be
estimated accurately almost surely. Redundancy is
one way to measure the ``frequentist'' properties of the Bayesian approaches 
we consider in this paper and refers to the compressibility of the distribution
from an information theoretic~perspective. 

As mentioned in the Introduction, redundancy differs from notions, such
as the concentration of the posterior distribution about the true
distribution. However, the parameters of the CRPs or the PY processes
that compress well often correspond to the maximum likelihood (ML) %
 estimates from the~sample.

\subsection{Weak Universality}

In the previous section, we considered guarantees that hold over the entire
model class; both the worst case and average case involve taking a supremum
over the entire model class. Therefore, the strong guarantees---average or
worst-case---hold uniformly over the model class. However, as we will see,
exchangeable estimators, in particular the Chinese restaurant process and Pitman--Yor estimators, are tuned towards specific kinds of sources, rather than all i.i.d.
models by the appropriate choice of parameters. This behavior is better captured by
looking at the model-dependent convergence of the exchangeable estimators, which is known as
weak universality. Specifically, let $\cP^\infty$ be a collection of i.i.d. measures
over infinite sequences, and let $\cP^\infty_\Psi$ be the collection of measures induced on patterns
by $\cP^\infty$.
We say an estimator $q$ is weakly universal for a class $\cP^\infty_\Ptr$ 
if for all $p\in\cP^\infty$:
	\begin{align*}
\limsup_{n\to\infty}\frac1n\mathbb{E}_p \left[ \log \frac{p(\ptr^n)}{q(\ptr^n)} 
\right]
=0.
	\end{align*}

\section{Strong Redundancy}
\label{sec:results}

We now describe our main results on the redundancy of estimators
derived from the prior distributions on $\mc{I}$. %

\subsection{Chinese Restaurant Process Predictors}

Previously \cite{SanthanamM:10isit}, it was shown by some of the authors that the
worst-case and average-case redundancies for the CRP estimator are both
$\Omega(n \log n)$, which means it is not strongly universal. However, this
negative result follows because the CRP estimator is tuned not to the entire
i.i.d. class of distributions, but to a specific subclass of i.i.d. sources
depending on the choice of parameter. To investigate this further, we analyze
the redundancy of the CRP estimator when we have a bound on the number $m$ of
distinct elements in the pattern $\ptr_1^n$.

Chinese restaurant processes $\pcrp$ with parameter $\theta$ are known
to generate exchangeable random partitions where the number of
distinct parts $M$ satisfy $M/\log n\to \theta$ almost surely as the
sample size $n$ increases; see \eg~\cite{Car99}. Equivalently, the CRP
generates patterns with $M$ distinct symbols, where $M/\log n\to
\theta$. However, the following theorem reverses the above setting.
Here, we are given an i.i.d. sample of data of length $n$ with $m$
symbols (how the data were generated is not important), but we
pick the parameter of a CRP estimator that describes the
pattern of the data well. While it is satisfying that the
chosen parameter matches the ML estimate of the number of symbols,
note that this need not necessarily be the only parameter choice that
works.

\bTheorem [Redundancy for CRP estimators]
\label{thm:crp:red}
Consider the estimator $\pcrp{\ptr_1^n}$ in \eqref{eq:CRPfull} and \eqref{eq:CRPstep}. Then, for sufficiently large $n$ and for patterns $\psi_1^n$ whose number of distinct symbols $m$ satisfies:
	\begin{align*}
	m \leq C \cdot \frac{n}{\log n} (\log \log n)^2,
	\end{align*}
the redundancy of the predictor $\pcrp{\ptr_1^n}$ with $\lceil \theta \rceil = m/\log n$ satisfies:
	\begin{align*}
	\log \frac{p(\ptr_1^n)}{\pcrp{\ptr_1^n}} \le 3C \cdot \frac{n (\log \log n)^3}{ \log n } = o(n). 
	\end{align*}
\Proof
The number of patterns with prevalence $\{\phi_{\mu}\}$ is:
	\begin{align*}
	\frac{n!}{\prod_{\mu=1}^{n} [\mu!]^{\phi_{\mu}} \phi_{\mu}! }, 
	\end{align*}
and therefore:
	\begin{align}
	p(\ptr_1^n) \le \frac{\prod_{\mu=1}^{n} [\mu!]^{\phi_{\mu}} \phi_{\mu}! }{n!},
	\label{eq:patprob}
	\end{align}
since patterns with prevalence $\{\phi_{\mu}\}$ all have the same probability.

Using the upper bound in \eqref{eq:patprob} on $p(\ptr_1^n)$ and \eqref{eq:CRPfull} yields:
	\begin{align}
	\log \frac{p(\ptr_1^n)}{\pcrp{\ptr_1^n}} 
		&\le
			\log \prod_{\mu=1}^{n} \frac{[\mu!]^{\phi_{\mu}} \phi_{\mu}!}{\left[(\mu - 1)!\right]^{\phi_{\mu}} \theta^m }
			+ \log \frac{ \theta (\theta + 1) \cdots (\theta + n - 1) }{ n!} \nonumber \\
		&= \log \left( \prod_{\mu=1}^{n} \mu^{\phi_{\mu}} \right) 
		+ \log \left( \frac{1}{\theta^m} \prod_{\mu=1}^{n} \phi_{\mu}! \right)
		+ \log \frac{ \theta (\theta + 1) \cdots (\theta + n - 1) }{ n!}.
		\label{eq:crp:ratio1}
	\end{align}
Let $\btheta = \lceil \theta \rceil$. The following bound follows from Stirling's approximation:
	\begin{equation}	
	\frac{ \theta (\theta + 1) \cdots (\theta + n - 1) }{ n!}
	\le \frac{ (\btheta + n)! }{ \theta! n! } 
	\le \frac{ (\btheta + n)^{\btheta + n} }{ \btheta^{\btheta} n^n }  
	\le \left( \frac{ \btheta + n }{\btheta} \right)^{\btheta}
		 \left( \frac{ \btheta + n }{n} \right)^{n}. \label{eq:crp:stirling}
	\end{equation}
The first term in \eqref{eq:crp:ratio1} can be upper bounded by $\log (n/m)^m$ since the argument of the $\log(\cdot)$ is maximized over $\mu \in [1,n]$ when $\mu = n/m$. The second term is also maximized when all symbols appear the same number of times, corresponding to $\phi_{\mu} = m$ for one $\mu$. Therefore:
	\begin{align*}
	\log \frac{p(\ptr_1^n)}{\pcrp{\ptr_1^n}} 
		&\le
		\log\left( \frac{n}{m} \right)^m + \log \frac{m!}{\theta^m} %
		+ \log \left( \frac{ \btheta + n }{\btheta} \right)^{\btheta}
		 \left( 1 + \frac{ \btheta}{n} \right)^{n}.
	\end{align*}
Now, $\left( 1 + \frac{ \btheta}{n} \right)^{n} \le e^{\btheta}$ for sufficiently large $n$, so: 
	\begin{equation}
	\log \frac{p(\ptr_1^n)}{\pcrp{\ptr_1^n}} 
		\le
		\log\left( \frac{n}{m} \right)^m + \log \frac{m!}{\theta^m} 
		+ \log \left( \frac{ (\btheta + n) e }{\btheta} \right)^{\btheta}.
	\label{eq:crp:stairbound}
	\end{equation}
Choose $\btheta = m/\log n$. This gives the bound:
	\begin{align*}
	\log \frac{p(\ptr_1^n)}{\pcrp{\ptr_1^n}} 
		&\le
		m \log\left( \frac{n}{m} \right) 
		+ \log \frac{m!}{m^m} \left( \frac{\btheta}{\theta} \right)^m
		+ m \log \log n
		+ \log \left( \frac{ (\btheta + n) e }{\btheta} \right)^{\btheta}.
	\end{align*}
the second term is negative for sufficiently large $m$. Therefore:
	\begin{align}
	\log \frac{p(\ptr_1^n)}{\pcrp{\ptr_1^n}} 
		&\le
		m \log\left( \frac{n}{m} \right) 
		+ m \log \log n
		+ \frac{m}{\log n} \log \left( 2 + \frac{n \log n}{m} \right).
	\label{eq:crp:partialred}
	\end{align}
Noting that the function above is monotonic in $m$ for $n\ge16$, we choose:
	\begin{align*}
	m = C \frac{n}{\log n} (\log \log n)^2,
	\end{align*}
and the bound becomes:
	\begin{align*}
	\log \frac{p(\ptr_1^n)}{\pcrp{\ptr_1^n}} 
		& \\
		& \hspace{-0.7in} \le
		C n \frac{(\log \log n)^2 }{\log n} \log\left( \frac{\log n}{ (\log \log n)^2 } \right)
		+ C n \frac{(\log \log n)^3}{\log n}
		+ C n \left(\frac{\log \log n}{\log n} \right)^2 \log \left( 2 + \left(\frac{\log n}{\log \log n}\right)^2 \right) \\
		& \hspace{-0.7in} \le
		3 C n \frac{(\log \log n)^3}{\log n} \\
		& \hspace{-0.7in} =
		o(n).\tag*{$\Box$}
	\end{align*}
\eTheoremp

This theorem is slightly dissatisfying, since it requires us to have a bound on
the number of distinct symbols in the pattern. In Section~\ref{s:wu}, we take
mixtures of CRP estimators to arrive at estimators that are weakly universal.

\subsection{Pitman--Yor Predictors}

We now turn to the more general class of Pitman--Yor predictors. 
We can obtain a similar result as for the CRP estimator, but we can handle 
all patterns with $m = o(n)$ distinct symbols.

As before, the context for the following theorem is this: we are given
an i.i.d. sample of data of length $n$ with $m$ symbols (there is no
consideration, as before, as to how the data was generated), but we
pick the parameters of a PY estimator that describes the
pattern of the data well. The choice of the PY estimator is not necessarily
the best, but one that will help us construct the weakly universal 
estimator in later sections of this paper.

We also note that the choice of the parameter $\theta$ below is
analogous to our choice when $\alpha=0$ (reducing to the CRP
case). For patterns generated by a PY process, where $0<\alpha\le 1$,
the number of distinct symbols grows as $n^\alpha$. It is known that
in this regime, the choice of $\theta$ is not
distinguishable~\cite{PY97}. However, what is known is that the choice
of $\theta$ remains $o(n^\alpha)$, something that is achieved in the
selection of $\theta$ below. As the reader will note, as long as
$0<\alpha<1$ is fixed, the theorem below will help us construct weakly
universal estimators further on.

\bTheorem
[Worst-case redundancy]
Consider the estimator $\ppy{\ptr_1^n}$. Then, for sufficiently large $n$ and for patterns $\psi_1^n$, whose number of distinct symbols $m$ satisfies $m = o(n)$, the worst-case redundancy of the predictor $\ppy{\ptr_1^n}$ with $\theta = m/\log n$ satisfies:
	\begin{align}
	\log\frac{p(\ptr_1^n)}{\ppy{\ptr_1^n}} = o(n). 
	\end{align}
\Proof
For a pattern $\ptr_1^n$, from the definition of $\ppy{\ptr_1^n}$ in \eqref{eq:PYfull} and \eqref{eq:patprob},
	\begin{align}
	\log\frac{p(\ptr_1^n)}{\ppy{\ptr_1^n}}
	&\le \log \left( 
		\frac{\prod_{\mu=1}^{n} [\mu!]^{\phi_{\mu}} \phi_{\mu}! }{n!} \cdot
		\frac{(\theta+1) \cdots (\theta + n - 1)}{(\theta + \alpha) (\theta + 2 \alpha) \cdots (\theta + m \alpha)}
		\prod_{\mu=1}^{n} \left( \frac{\Gamma(1 - \alpha)}{\Gamma(\mu - \alpha - 1)} \right)^{\phi_{\mu}}
		\right). %
	\end{align}
We can bound the components separately. First, as before we have:
	\begin{align*}
	\prod_{\mu=1}^{n} \phi_{\mu}! \le m!
	\end{align*}	
Since $\theta > -\alpha$, we have $\theta + \alpha > 0$ and:
	\begin{align*}
	(\theta + \alpha) (\theta + 2 \alpha) \cdots (\theta + (m-1) \alpha)
	&\ge (\theta + \alpha) \alpha (2 \alpha) \cdots ((m-2) \alpha) \\
	&= (\theta + \alpha) (m-2)! \alpha^{m-2}.
	\end{align*}
Again, letting $\btheta = \lceil \theta \rceil$, from the same arguments as in \eqref{eq:crp:stirling} and \eqref{eq:crp:stairbound},
	\begin{align*}
	\log \frac{\theta (\theta+1) \cdots (\theta + n - 1)}{ n! } 
	\le \btheta \log \frac{ (\btheta + n) e }{\btheta}.
	\end{align*}
Finally, note that $(1 - \alpha) (2 - \alpha) \cdots (\mu - 1 - \alpha) \ge (1 - \alpha) (\mu-2)!$, so:
	\begin{align*}
	\frac{
		\prod_{\mu=1}^{n} [\mu!]^{\phi_{\mu}}
		}{ 
		\prod_{\mu=1}^{n} [(1 - \alpha) (2 - \alpha) \cdots (\mu - 1 - \alpha) ]^{\phi_{\mu}} 
		}
	&\le \prod_{\mu=1}^{n} \left( \frac{ \mu! }{ (1 - \alpha) (\mu - 2)! } \right)^{\phi_{\mu}} \\
	&\le \frac{ \prod_{\mu=1}^{n} \mu^{2 \phi_{\mu}} }{ (1 - \alpha)^{m} } \\
	&\le \frac{ (n/m)^{2m} }{ (1 - \alpha)^{m} }.
	\end{align*}

Putting this together:
	\begin{align}
	\log\frac{p(\ptr_1^n)}{\ppy{\ptr_1^n}}
	&\le
	\log\frac{ m! }{ (\theta + \alpha) (m-2)! \alpha^{m-2} }
		+ \btheta \log \frac{ (\btheta + n) e }{\btheta}
		+ \log \frac{ (n/m)^{2m} }{ (1 - \alpha)^{m} } \nonumber \\
	&\le 2m \log \frac{n}{m} 
		+ (m-2) \log \frac{1}{(1 - \alpha) \alpha}
		+ \btheta \log \frac{ (\btheta + n) e }{\btheta}
		+ \log \frac{ m^2 }{ (\theta + \alpha) }
		+ \log \frac{1}{(1 - \alpha)^2}.
	\label{eq:PYupper}
	\end{align}
If $m = o(n)$, then the right side above is less than $o(n)$, as desired.
\eTheorem

It is well known that the Pitman--Yor process can produce patterns
whose relative frequency is zero, e.g. the pattern $1^k 2 3 \cdots
(n-k)$. Therefore, it is not surprising that the worst-case redundancy
and average case redundancies can be bad. However, as the next
theorem shows, the actual redundancy of the Pitman--Yor estimator is
$\Theta(n)$, which is significantly better than the lower bound of
$\Omega(n\log n)$ proven in Santhanam and
Madiman~\cite{SanthanamM:10isit} for Chinese restaurant processes.

\bTheorem
[Redundancies]
Consider the estimator $\ppy{\ptr_1^n}$. Then, for sufficiently large $n$, the worst-case redundancy and average case redundancy satisfy:
	\begin{align}
	\hat{R}(\ppy{\cdot}) = \Theta(n) \qquad \textrm{and} \qquad
	\bar{R}(\ppy{\cdot}) = \Theta(n).
	\end{align}
That is, $\ppy{\cdot}$ is neither strongly nor weakly universal.
\Proof
For the upper bound, we start with \eqref{eq:PYupper} and note that in the worst case, $m = O(n)$, so $\hat{R}(\ppy{\cdot}) = O(n)$ and \textit{a fortiori} $\bar{R}(\ppy{\cdot}) = O(n)$.

For the lower bound, consider the patterns $1 1 \cdots 1$ and $1 2 \cdots n$. For the Pitman--Yor estimator,
	\begin{align*}
	\ppy{1 1 \cdots 1} \ppy{1 2 \cdots n}
	&= \frac{\theta (1 - \alpha) \cdots (n-1 - \alpha) }{ \theta (\theta + 1) \cdots (\theta + n-1)}
		\frac{\theta (\theta + \alpha) \cdots (\theta + (n-1)\alpha) }{ \theta (\theta + 1) \cdots (\theta + n-1)} \\
	&= \frac{ (1 - \alpha) ( \theta + \alpha) }{ (\theta + 1)^2 }
		\cdot 
		\frac{ (2 - \alpha) ( \theta + 2 \alpha) }{ (\theta + 2)^2 } 
		\cdots
		\frac{ (n-1 - \alpha) ( \theta + (n-1) \alpha) }{ (\theta + n-1)^2 }.
	\end{align*}

For $j \ge 1$, $0<\alpha<1$ and $\alpha+\theta >0$, we show in the claim proven below that: 
\[
	\frac{(j - \alpha)(\theta + j \alpha)}{ (\theta + j)^2 } 	\le \max\{ \frac{1}{2}, \alpha \}.
\]
 Therefore, each term is less than one. Then, for $\alpha < 1$, there exists a constant $0 < c < 1$, such that:
	\begin{align*}
	\ppy{1 1 \cdots 1} \ppy{1 2 \cdots n} \le c^n.
	\end{align*}
Thus: 
	\begin{align*}
	\log \frac{1}{\ppy{1 1 \cdots 1}} + \log \frac{1}{ \ppy{1 2 \cdots n} } \ge n \log \frac{1}{c}.
	\end{align*}
Let the distribution $p_1$ be a singleton, so $p_1(1 \cdots 1) = 1$. For any small $\delta > 0$, we can find a distribution $p_n$, such that $p_n(1 2 \cdots n) = 1 - \delta$ by choosing $p_n$ to be uniform on a sufficiently large set. Thus:
	\begin{align*}
	\hat{R}(\ppy{\cdot}) 
	&\ge \max \left\{ \log \frac{1 - \delta}{\ppy{1 1 \cdots 1}}, \log \frac{1 - \delta}{ \ppy{1 2 \cdots n}} \right\} \\
	&\ge \frac{1}{2} \left( \log \frac{1}{\ppy{1 1 \cdots 1}} + \log \frac{1}{ \ppy{1 2 \cdots n} } \right) + \log(1 - \delta) \\
	&\ge \frac{n}{2} \log \frac{1}{c} + \log(1 - \delta).
	\end{align*}
This shows that $\hat{R}(\ppy{\cdot}) = \Omega(n)$. Furthermore,
	\begin{align*}
	\bar{R}(\ppy{\cdot}) 
	&\ge \max \left\{ (1 - \delta) \frac{1 - \delta}{\ppy{1 1 \cdots 1}}, (1 - \delta) \frac{1 - \delta}{ \ppy{1 2 \cdots n} } \right\} \\
	&\ge (1 - \delta) \left( \frac{n}{2} \log \frac{1}{c} + \log (1 - \delta) \right),
	\end{align*}
so $\bar{R}(\ppy{\cdot}) = \Omega(n)$.

All that remains is to prove the following claim:
\bClaim
For $j \ge 1$, $0<\alpha<1$ and $\alpha+\theta >0$ we show that: 
\[
	\frac{(j - \alpha)(\theta + j \alpha)}{ (\theta + j)^2 } 	\le \max\{ \frac{1}{2}, \alpha \}.
\]
\Proof
	First, assume that $0<\alpha< \frac{1}{2}$. Then, the inequality is:
		\begin{align*}
		\frac{(j - \alpha)(\theta + j \alpha)}{ (\theta + j)^2 } \le \frac{1}{2}.
		\end{align*}
	Equivalently, we need to show:
		\begin{align*}
		(1-2\alpha)j^2 +2\alpha^2 j+\theta^2+2\alpha\theta \geq 0.
		\end{align*}
	Since $1-2\alpha > 0$, the quadratic is always nondecreasing on $j\geq 1$. Therefore, the positive integer $j=1$
	minimizes the quadratic above, and the value of the quadratic at $j=1$ is:
		\begin{align*}
		1-2\alpha +2\alpha^2 +\theta^2+2\alpha\theta = (\alpha - 1)^2 + (\theta + \alpha)^2 \geq 0.
		\end{align*}
	Next, assume that $\frac{1}{2}\le\alpha<1$. Then, the inequality is:
		\begin{align*}
		\frac{(j - \alpha)(\theta + j \alpha)}{ (\theta + j)^2 } \le \alpha.
		\end{align*}
	Equivalently, we need to show:
		\begin{align}\label{claim_equation}
		((2\alpha - 1)\theta + \alpha^2)j +\alpha\theta(\theta+1) \geq 0.
		\end{align}
	Since $2\alpha - 1 \ge 0$ and $\theta>-\alpha$,
		\begin{align*}
		(2\alpha - 1)\theta + \alpha^2 \ge -(2\alpha - 1)\alpha + \alpha^2=\alpha(1-\alpha) > 0.
		\end{align*}
	Therefore, the minimum of the left equation in (\ref{claim_equation}) is achieved at $j=1$. Note that $\alpha\theta^2 > -\alpha^3 -2\alpha^2\theta$ by using $(\alpha+\theta)^2>0$. Therefore, the value of the minimum is:
		\begin{align*}
		(2\alpha - 1)\theta + \alpha^2 +\alpha\theta(\theta+1) &= \alpha\theta^2 + (3\alpha -1)\theta + \alpha^2 \geq -\alpha^3 + (-2\alpha^2 + 3\alpha -1)\theta + \alpha^2 \\
		&\geq -\alpha^3 -(-2\alpha^2 + 3\alpha-1)\alpha + \alpha^2 \\
		&=\alpha^3-2\alpha^2 +\alpha = \alpha(\alpha-1)^2\geq 0.
		\end{align*}
	Note that $-2\alpha^2 + 3\alpha -1 \geq 0$ for $\frac{1}{2}\le\alpha<1$, and the claim follows.
\eClaim
The theorem follows.
\eTheorem

\section{Weak Universality}
\label{s:wu}

In this section, we show how to modify the CRP or PY estimators to obtain weakly universal
estimators. The CRP and PY cases are identical; therefore, we only work out the CRP case.

For all $i\ge1$ and $j\ge1$, let:
\begin{align*}
 c_{i,j} = \frac{1}{i (i+1) j (j+1)}
\end{align*}
so that $\sum_{i,j} c_{i,j} = 1$. Let $\pcrpmixij{\cdot}$ be the CRP measure
over patterns with $\theta = i/\log j$. Consider the following measure over
patterns of infinite sequences that assigns, for all $n$ and all patterns
$\ptr^n$ of length $n$, the probability:
\begin{align}
 q^{\ast}(\ptr^n) = \sum_{i,j} c_{i,j} \pcrpmixij{\ptr^n}.
 \label{eq:crp_mixture}
\end{align}
We will show that $q^{\ast}$ is a weakly universal measure over patterns of i.i.d.
sequences.

To do so, we will need the following two lemmas. Lemma~\ref{lem:mnk} is a useful ``folk'' inequality
that we believe is attributed to Minkowski. Lemma~\ref{lem:Hrate} relates the expected number of 
distinct symbols in length $n$ sequences of an i.i.d. process to its entropy and
is of independent interest. The result not only strengthens a similar result 
in~\cite{OSVZ04:lim}, but also provides a different and more compact proof.

\bLemma
\label{lem:mnk}
For $n\ge 1$, let $x_1\ge x_2\ge\ldots x_n\ge 0$ and $y_1\ge y_2 \ge\ldots y_n\ge 0$ be two 
sorted sequences. Then:
\[
\frac1n \sum_{l=1}^n x_l y_l 
\ge 
\Paren{\frac1n\sum_{i=1}^n x_i}\Paren{\frac1n\sum_{j=1}^n y_j} 
\ge
\frac1n \sum_{l=1}^n x_l y_{n+1-l}.
\]
\Proof
The left inequality of the lemma
follows by noting that:
\[
\Paren{\sum_{i=1}^n x_i}\Paren{\sum_{j=1}^n y_j} 
=
\sum_{k=0}^{n-1} \sum_{l=1}^n x_l y_{l+k},
\]
and that the sum $\sum_{l} x_l y_{l+k}$ is maximized at $k=0$, since both sequences are sorted
in the same direction. The right inequality of the lemma can be proven similarly, but will not be
used in the paper.
\eLemma
\bLemma
\label{lem:Hrate}
For all discrete i.i.d. processes $P$ with entropy rate (or marginal entropy) $H$, let $M_n$ be the random variable 
counting the number of distinct symbols in a sample of length $n$ drawn from $P$. The following bound holds:
	\begin{align}
	\mathbb{E}\left[ M_n \right] \le \frac{n H}{\log n} + 1.
	\end{align}
\Proof
Let $P(i) = p_i$. We begin by noting that: 
\[
H= \sum_i p_i \log \frac1{p_i}
=
\sum_i p_i \sum_{j=1}^\infty \frac{(1-p_i)^j}j,
\]
where the second equality follows by the Taylor series expansion: 
\[
-\log p_i= -\log (1-(1-p_i)) = \sum_{j=1}^\infty \frac{(1-p_i)^j}j.
\]
The right summation in the equation above is bounded below as follows:
\begin{align*}
\sum_{j=1}^\infty \frac{(1-p_i)^j}j 
&\ge
\sum_{j=1}^n \frac{(1-p_i)^j}j \\
&\age{(a)}
\frac1n 
\sum_{l=1}^n \frac1l 
\sum_{m=1}^n (1-p_i)^m \\
&\ge
\frac{\log n}n
\frac{(1-p_i)}{p_i}\Paren{1-(1-p_i)^n}
\end{align*}
where $(a)$ follows from Minkowski's inequality in Lemma~\ref{lem:mnk} and the last inequality,
because $\sum_{l=1}^n \frac1l \ge \log n$.
Thus,
\[
H= 
\sum_i p_i \sum_{j=1}^\infty \frac{(1-p_i)^j}j
\ge
\frac{\log n}{n}
\sum_i (1-p_i)
\Paren{1-(1-p_i)^n} 
\ge
\frac{\log n}{n} \Paren{\E M_n - 1}
\]
where for the second inequality, we use $\sum_{i} p_i \Paren{1-(1-p_i)^n} \le \sum_{i} p_i \le 1$.
\ignore{
Writing out the expectation of $M_n$, we see:
	\begin{align*}
	\mathbb{E}\left[ M_n \right] &= \sum_{i=1}^{\infty} (1 - (1 - p_i)^n) \\
	&= \sum_{i=1}^{\infty} p_i \frac{ 1 - (1 - p_i)^n }{1 - (1 - p_i) } \\
	&= \sum_{i=1}^{\infty} p_i \sum_{j=0}^{n-1} (1 - p_i)^{j} \\
	&\le 1 + \sum_{i=1}^{\infty} p_i \sum_{j=1}^{n-1} (1 - p_i)^{j}.
	\end{align*}
Now, using the fact that $\sum_{k=1}^{n-1} \frac{1}{k} \ge \log n$,
	\begin{align*}
	\mathbb{E}\left[ M_n \right] \log n
	&\le \log n + \sum_{i=1}^{\infty} p_i \left( \sum_{j=1}^{n-1} (1 - p_i)^{j} \right)
		\left( \sum_{k=1}^{n-1} \frac{1}{k} \right) \\
	&= \log n + \sum_{i=1}^{\infty} p_i 
		\sum_{j=1}^{n-1} 
		\sum_{k=1}^{n-1} 
			\frac{(1 - p_i)^{j}}{k} \\
	&\le \log n + n \sum_{i=1}^{\infty} p_i \sum_{j=1}^{n-1} \frac{(1 - p_i)^{j}}{j} \\
	&\le \log n + n \sum_{i=1}^{\infty} p_i \sum_{j=1}^{\infty} \frac{(1 - p_i)^{j}}{j} \\
	&= \log n + n \sum_{i=1}^{\infty} - p_i \log(1 - (1 - p_i)) \\
	&= \log n + n H.
	\end{align*}}
\eLemma

\bTheorem
[Weak universality for CRP mixtures]
For all discrete i.i.d. processes $p \in \cI$ with a finite entropy rate,
	\begin{align}
	D\left( p ~\|~ q^{\ast} \right) = o(n).
	\end{align}
That is, $q^{\ast}$ is weakly universal.
\Proof
We write the divergence between $p$ and $q^{\ast}$ in \eqref{eq:crp_mixture} as the expected log ratio and condition on the value of $M_n$:
	\begin{eqnarray}
	&&\mathbb{E}_p\left[ \log \frac{ p(\ptr_1^n) }{ \sum_{m} c_{m,n} \pcrpmix{\ptr_1^n} } \right] \nonumber \\
	&= &\mathbb{P}\left( M_n > \frac{n (\log \log n)^2}{\log n} \right) \cdot
		\mathbb{E}_p\left[ \log \frac{ p(\ptr_1^n) }{ \sum_{m} c_{m,n} \pcrpmix{\ptr_1^n} } 
			\Big|
			M_n > \frac{n (\log \log n)^2}{\log n}
			\right] \nonumber \\
		&&+\mathbb{P}\left( M_n < \frac{n (\log \log n)^2}{\log n} \right) \cdot
		\mathbb{E}_p\left[ \log \frac{ p(\ptr_1^n) }{ \sum_{m} c_{m,n} \pcrpmix{\ptr_1^n} } 
			\Big|
			M_n < \frac{n (\log \log n)^2}{\log n}
			\right].
		\label{eq:crpmix_div}
	\end{eqnarray}

Consider the estimator $\pcrpmixij{\ptr_1^n}$ in $q^{\ast}$ corresponding to $i = M_n$ and $j = \log n$. This is the estimator $\pcrp{\ptr_1^n}$ with $\theta = M_n/log n$. From the proof of Theorem \ref{thm:crp:red}, we have:
	\begin{align}
	\log \frac{ p(\ptr_1^n) }{ \sum_{m} c_{m,n} \pcrpmix{\ptr_1^n} }
	&\le \log \frac{p(\ptr_1^n)}{ c_{i,j} \pcrpmixij{\ptr_1^n} } \nonumber \\
	&\le \log \frac{1}{c_{ij}} + \log \frac{p(\ptr_1^n)}{ \pcrpmixij{\ptr_1^n} } \nonumber \\
	&\le \log\left( M_n (M_n + 1) (\log n)( \log n + 1) \right) + \log \frac{p(\ptr_1^n)}{ \pcrp{\ptr_1^n} }
		\label{eq:crpmix_bound}
	\end{align}
We will bound the two terms in \eqref{eq:crpmix_bound} in the regimes for $M_n$.	

The result of Theorem \ref{thm:crp:red} says that if $M_n < \frac{n (\log \log n)^2}{\log n}$, 
then:
	\begin{align}
	\log \frac{1}{c_{ij}} + \log \frac{p(\ptr_1^n)}{ \pcrp{\ptr_1^n} }
	&\le \log\left( M_n (M_n + 1) (\log n)( \log n + 1) \right) + o(n)
		\label{eq:smallM_n}
	\end{align}
Thus, this term is $o(n)$.

If $M_n > \frac{n (\log \log n)^2}{\log n}$, then we first apply Markov's inequality using the previous lemma:
	\begin{align}
	\mathbb{P}\left( M_n > \frac{n (\log \log n)^2}{\log n} \right)
	&\le \frac{\log n}{n (\log \log n)^2} \left( \frac{n H}{\log n} + 1 \right) \nonumber \\
	&\le \frac{H}{(\log \log n)^2} + \frac{\log n}{n (\log \log n)^2}.
	\label{eq:distinct:markov}
	\end{align}
Therefore, for all finite entropy processes, this probability goes to zero as $n \to \infty$. Looking at the term in $q^{\ast}$ corresponding $\pcrp{\ptr_1^n}$ with $\theta = M_n/\log n$ and using the fact that $M_n \le n$, we see that the first term in \eqref{eq:crpmix_bound} is upper bounded as $O(\log n)$. For the second term, we appeal to \eqref{eq:crp:partialred} in the proof of Theorem~\ref{thm:crp:red}:
	\begin{align}
	\log \frac{p(\ptr_1^n)}{ \pcrp{\ptr_1^n} } 
	&\le M_n \log \frac{ M_n }{ n } + M_n \log \log n + \frac{M_n}{\log n} \log\left(2 + \frac{n \log n}{M_n} \right) \\
	&\le n + n \log \log n + n \frac{ \log\left(2 + n \log n \right) }{ \log n } \\
	&\le 3 n \log \log n.
	\end{align}
	
Plugging these terms into \eqref{eq:crpmix_div}:
	\begin{align*}
	D\left( p ~\|~ q^{\ast} \right) 
	&\le \left( \frac{H}{(\log \log n)^2} + \frac{\log n}{n (\log \log n)^2} \right) \cdot O(n \log \log n) 
		+ 1 \cdot o(n) = o(n).\tag*{$\Box$}
	\end{align*}
\eTheoremp

The preceding theorem shows that the mixture of CRP estimators $q^{\ast}$ is
weakly universal. However, note that $q^{\ast}$ is not itself a CRP estimator. An identical
construction is possible for the PY estimators, as well.
The convergence of the weakly universal $q^\ast$ depends on the number of entropy of the
source, as well as the number of distinct symbols in a sample of size $n$. 

While it would be tempting to predict the performance of the estimator $q^\ast$ for larger
sample sizes $N\ge n$, such a task requires a more careful analysis. In general, it may
be impossible to non-trivially bound the number of distinct symbols $M_N$ with 
smaller sample size $n$, as the following example shows.

\bExample Let $n=\sqrt{N}$. Consider a set $\cI$ containing the following two
distributions: (i) $p$ over $\naturals$, which assigns probability $1-1/n^{3/2}=1-1/N^{3/4}$
to the atom $1$ and splitting %
 the probability $1/N^{3/4}$ equally among the elements of the set
$\sets{2\upto N^2+1}$; and (ii) $p'$, which simply assigns probability one to one. A
sample of size $n$ from either $p$ or $p'$ is $1^n$ with probability at least
$1-1/N^{1/4}$, no matter what the underlying source is; therefore, we cannot
distinguish between these sources with probability $1-1/N^{1/4}$ from a sample
of size $n$.

However, a sample of size $N$ from $p$ has $\cO(N^{1/4})$ distinct symbols on
average, while that of $p'$ will have only one element. It follows that if all we
know is that the unknown distribution comes from $\cI$, with high probability %
 under the unknown
source, we cannot predict whether the number of symbols in a sample of size $N$
will remain one or not from a sample of length $n$. Furthermore, by changing the
ratio of $n$ and $N$ (and therefore, the probability of the symbol 1 under $p$),
we can make the expected number of symbols in a $N-$length sample under $p$ as
large as we want. \eExample

However, it is possible to impose restrictions on the class of distributions that
allow us to ensure that we can predict the number of symbols in longer samples. 
In future work, we will borrow from the data-derived consistency formulations of \cite{SAKS14:isit} to characterize when we will be able to predict the number of symbols in longer samples.

\section{Conclusions and Future Work}

In this note, we investigated the worst-case and average-case
redundancies of pattern probability estimators derived from priors on
$\mc{I}$ that are popular in Bayesian statistics. Both the CRP and
Pitman--Yor estimators give a vanishing redundancy per symbol for patterns whose number of
distinct symbols $m$ is sufficiently small. The Pitman--Yor estimator requires
only that $m = o(n)$, which is an improvement on the CRP. However,
when $m$ can be arbitrarily large (or the alphabet size is arbitrarily
large), the worst-case and average-case redundancies do not scale like
$o(n)$. Here, again, the Pitman--Yor estimator is superior, in that the
redundancies scale like $\Theta(n)$ as opposed to the $\Omega(n \log n)$ 
for the CRP estimator. 
While these results show that these estimators are not strongly universal,
we constructed a mixture of CRP process (which is not itself a CRP estimator) 
that is weakly universal. One of the estimators derived in~\cite{OSZ03} is
exchangeable and has near-optimal worst case redundancy of $O(\sqrt{n})$. 
Kingman's results imply this estimator corresponds to a prior on $\cI$; 
however, this prior is yet unknown. Finding
this prior may potentially reveal new interesting classes of priors
other than the Poisson--Dirichlet priors.

\section*{Acknowledgments}

The authors thank the American Institute of Mathematics and NSF for
sponsoring a workshop on probability estimation, as well as A.~Orlitsky 
and K.~Viswanathan, who co-organized the workshop with the first author.
They additionally thank P.~Diaconis, M.~Dudik, F.~Chung,
R.~Graham, S.~Holmes, O.~Milenkovic, 
O.~Shayevitz, A.~Wagner, J.~Zhang and M.~Madiman for 
helpful discussions.

N.P.~Santhanam was supported by a startup grant from the University of Hawaii and NSF Grants
CCF-1018984 and CCF-1065632.
A.D.~Sarwate was supported in part by
the California Institute for Telecommunications and Information Technology (CALIT2) at the University of California, San Diego.
J.O.~Woo was supported by NSF Grant CCF-1065494 and CCF-1346564.

\bibliographystyle{ieeetr}
\bibliography{univcod2}

\end{document}